\documentclass[nato,numreferences]{crckbked}
\input amssym.def
\begin{document}

\begin{article}
\begin{opening}

\title{ADELIC QUANTUM MECHANICS:\\ NONARCHIMEDEAN AND NONCOMMUTATIVE ASPECTS}

\author{Goran \surname{Djordjevi\'c$^{1,2}$, Branko Dragovich$^{3,4}$, Ljubi\v sa Ne\v si\'c$^1$}
\thanks{\email{gorandj@junis.ni.ac.yu}}}

\institute{$^1$ Department of Physics, Faculty of Sciences, University of Ni\v
s, P.O. Box 91, 18001 Ni\v s, Yugoslavia}
\institute{$^2$ Sektion Physik, Universit\"at M\"unchen, Theresienstr. 37,
D-80333 M\"unchen, Germany}
\institute{$^3$ Institute of Physics, P.O.Box 57, 11001 Belgrade, Yugoslavia}
\institute{$^4$ Steklov Mathematical Institute, Gubkin St. 8, 117966,
  Moscow, Russia}

\begin{abstract}

We present a short review of adelic quantum mechanics pointing out
its non-Archimedean and noncommutative aspects. In particular,
$p$-adic path integral and adelic quantum cosmology are considered.
Some similarities between $p$-adic analysis  and q-analysis are noted. 
The $p$-adic Moyal product is introduced.

\end{abstract}
\end{opening}

\section{Introduction}

There is now a common belief that the usual picture of spacetime as a smooth
pseudo-Riemannian manifold should breakdown somehow at the Planck length
$l_p\sim 10^{-33}cm$,
due to the  quantum gravity effects. 
We consider here two possibilities, which come from 
modern mathematics  and mathematical physics: 
non-Archimedean geometry related to  $p$-adic numbers, and noncommutative geometry
with  space coordinates given by noncommuting operators
\begin{equation}
\label{djo-noncom}
[\hat x^i,\hat x^j]=i{\hbar}\theta^{ij}
\end{equation}
or by q-deformation $x^ix^j=qx^jx^i$.
Some noncommutativity of configuration space should not be a surprise in
physics since quantum phase space with the canonical commutation relation
(\ref{djo-com})
is the well-known example of noncommutative geometry.
We will mostly review our recent results concerning adelic
quantum mechanics.
We illustrate some features of adelic quantum mechanics by its application in quantum
cosmology. A few remarkable similarities between 
non-Archimedean and   noncommutative structures
are noted. The usual Moyal product is extended to $p$-adic and adelic quantum mechanics.

Since 1987, there have been many interesting applications of $p$-adic
numbers and non-Archimedean geometry in various parts of modern theoretical
and mathematical physics (for a review, see \cite{djo-breke,djo-vladi,djo-khren}). 
However we restrict ourselves here to $p$-adic
and adelic quantum mechanics as well as to some related topics. In particular, we review  Feynman's 
$p$-adic path integral method. A fundamental role of integral approach to $p$-adic and adelic
quantum mechanics (and adelic quantum cosmology) is emphasized.
The obtained $p$-adic probability amplitude for one-dimensional systems with quadratic
Lagrangians has  the form as that one in ordinary
quantum mechanics. 

It is well known that measurements give rational numbers $\Bbb Q$, whereas
theoretical models traditionally use real $\Bbb R$ and complex $\Bbb C$ number 
fields. A completion of $\Bbb Q$ with respect to the $p$-adic norms gives 
the fields of $p$-adic numbers $\Bbb Q_p$ ($p$ is a prime number) in the same way
as completion with absolute value yields $\Bbb R$.
The paper of Volovich \cite{djo-volo} 
initiated a series of articles on  $p$-adic string theory and many
other branches of theoretical and mathematical physics. The metric introduced 
by $p$-adic norm is the non-Archimedean (ultrametric) one. Possible existence of 
such space around the
Planck length is the main motivation to study
$p$-adic quantum models. However, $p$-adic analysis also plays a role in some areas of
"macroscopic physics" as, for example: spin glasses, quasicrystals and some other complex systems.

In order to investigate possible $p$-adic quantum phenomena it is necessary to have the
corresponding theoretical formalism.
An important step in this direction is a formulation  of 
$p$-adic quantum mechanics \cite{djo-vladi2,djo-ruelle}. Because of total 
disconnectedness of $p$-adic spaces and different
valuations of variables and wave functions, the quantization is 
performed by 
the Weyl procedure.  
A unitary
representation of the evolution operator $U_p(t)$ on the Hilbert space
${\cal L}_2(\Bbb Q_p)$ of complex-valued functions of a p-adic argument 
is an appropriate way to describe quantum dynamics
of $p$-adic systems. Recently formulated adelic 
quantum mechanics \cite{djo-mentor} successfully  unifies
ordinary and all $p$-adic quantum mechanics. The
appearance of space-time discreteness in adelic formalism 
 (see, e.g. \cite{djo-mentormi}) is an encouragement for the further 
investigations.

This paper is organized as follows. We start with a short introduction 
to $p$-adic numbers, adeles and their functions. After that, $p$-adic
and adelic quantum mechanics based on the Weyl quantization and Feynman's
path integral are presented. In Section 4 we review our
previuos results concerning one-dimensional $p$-adic propagator. In 
Section 5 we will see how adelic quantum
mechanics can be useful  in investigation of the very early universe, where
in a natural way space-time discreteness emerges in minisuperspace
models of adelic quantum cosmology. In the last Section we give some of interesting
relations between non-Archimedean and noncommutative analyses. We also define
and discuss the corresponding $p$-adic Moyal product.

\section{$p$-Adic numbers and adeles}

Any $x\in \Bbb Q_p$ can be presented in the form \cite{djo-shikhof}
\begin{equation}
\label{djo-canonexp}
x = p^\nu(x_0+x_1p+x_2p^2+\cdots)\ ,\quad \nu\in \Bbb Z,
\end{equation}
where $x_i = 0,1,\cdots,p-1$ are digits. $p$-Adic norm of any term
$x_ip^{\nu+i}$ in the canonical expansion (\ref{djo-canonexp}) is $\mid x_ip^{\nu+i}\mid_p
=p^{-(\nu+i)}$ and the strong triangle inequality 
holds,
 {\it i.e.} $\mid a+b\mid_p\leq\hbox{max}\{\mid
a\mid_p,\mid b\mid_p\}$. It follows that $\mid x\mid_p = p^{-\nu}$ if
$x_0\neq 0$. 
There is 
no natural ordering on $\Bbb Q_p$. However one can
introduce a linear order on $\Bbb Q_p$ by the following
definition: $x<y$ if $\mid x\mid_p<\mid y\mid_p$ or when 
$\mid x\mid_p = \mid y\mid_p$ there exists such index $m\geq0$ that
digits satisfy $x_0 = y_0, x_1 = y_1, \cdots,x_{m-1} = y_{m-1}\ 
,x_m<y_m$.

Derivatives of $p$-adic valued functions $\varphi : \Bbb Q_p\to \Bbb Q_p$ are 
defined as in the real case, but with respect to the
$p$-adic norm. There is no integral $\int\varphi(x)dx$ in a sense of
the Lebesgue measure \cite{djo-vladi}, but one can introduce 
$\int^b_a\varphi(x)dx
= \Phi(b)-\Phi(a)$ as a functional of analytic functions
$\varphi(x)$, where $\Phi(x)$ is an antiderivative of $\varphi(x)$. 
In
the case of map $f:\Bbb Q_p\to \Bbb C$ there is
well-defined Haar measure. We use here the Gauss integral
\begin{equation}
\int_{\Bbb Q_\upsilon}\chi_\upsilon(ax^2+bx)dx =
\lambda_\upsilon(a)\mid2a\mid^{-{1\over2}}_\upsilon{\chi}_v
\big(-{b^2\over4a}\big)\ ,\qquad a\not=0,
\end{equation}
where index $\upsilon$ denotes real ($\upsilon =\infty$) and $p$-adic cases, {\it 
i.e.} $\upsilon =\infty,2,3,5,\cdots\!$.
$\chi_\upsilon$ is an additive character: $\chi_\infty(x)=\exp(-2\pi ix)$,
$\chi_p(x) = \exp(2\pi i\{x\}_p)$,\quad  where $\{x\}_p$
is the fractional part of $x\in \Bbb Q_p$. $\lambda_\upsilon (a)$ is 
the complex-valued arithmetic function \cite{djo-vladi}.

An adele \cite{djo-geljfand} is an infinite sequence $a=(a_\infty, 
a_2,..., a_p,...)$, where
$a_\infty\in \Bbb R\equiv \Bbb Q_\infty$, $a_{p}\in \Bbb Q_{p}$ with a restriction 
that $a_{p} \in \Bbb Z_{p}$ for all but a finite set $S$ of primes
$p$. The set of all adeles $\Bbb A$  may be  regarded as a subset of
direct topological product $\Bbb Q_\infty\times\prod_p \Bbb Q_p$
whose  elements satisfy the above restriction, {\it i.e.}
\begin{equation}
\Bbb A=\cup_S \Bbb A(S), \qquad \Bbb A(S)=\Bbb R\times\prod_{p\in S}
\Bbb Q_p\times \prod_{p\notin S} \Bbb Z_p.
\end{equation}
$\Bbb A$ is a topological space, and can be considered as a ring with
respect to the componentwise addition and multiplication. An elementary function on  
adelic ring $\Bbb A$ is
\begin{equation}
\label{djo-elementary}
        \varphi (x)=\varphi _{\infty}(x_{\infty})\prod_{p}^{}\varphi 
_{p}(x_{p})
        =\prod_{v}^{} \varphi _{v}(x_{v})  \;\;
\end{equation}
with the main restriction that
$\varphi (x)$ must satisfy $\varphi _{p}(x_{p})=\Omega 
(|x_{p}|_{p})$ 
for all but a finite number of $p$, where
\begin{equation}
\label{djo-omega}
\Omega (\mid x\mid_p) = \left \{ \begin{array}{ll}
  1, & 0 \leq \mid x\mid_p \leq 1  ,\\
0, & \mid x\mid_p>1, \end{array} \right.
\end{equation}
is a characteristic function on the set of
$p$-adic integers $\Bbb Z_p= \{x \in \Bbb Q_p:|x|_{p}\leq 1\}$.
It should be noted that the Fourier transform of the characteristic
function (vacuum state) $\Omega(|x_p|)$ is $\Omega(|k_p|)$.

All finite linear combinations of elementary functions
(\ref{djo-elementary}) make  the set $\cal D(\Bbb A)$ of the Schwartz-Bruhat
functions. The Fourier transform of $\varphi(x)\in\cal D(\Bbb A)$ (that maps
$\cal D(\Bbb A)$ onto $\cal D(\Bbb A)$) is
\begin{equation}
\tilde \varphi(y)=\int_{\Bbb A}\varphi(x)\chi(xy)dx=
\int_\Bbb R\varphi_\infty(x)\chi_\infty(x y)dx
\prod_p\int_{\Bbb Q_p} \varphi_p(x)\chi_p(xy)dx,
\end{equation}
where $dx=dx_\infty dx_2\dots dx_p\dots$ is the Haar measure on $\Bbb A$.
The Hilbert space $L_2(\Bbb A)$ is a space of complex-valued functions
$\psi_1(x), \psi_2(x),\dots$, with the scalar product and norm
\begin{equation}
(\psi_1,\psi_2)=\int_\Bbb A \bar\psi_1(x)\psi_2(x)dx, \quad
||\psi||=(\psi,\psi)^{1/2}<\infty .
\end{equation}
A basis of the above space may be given by the
orthonormal eigenfunctions of  an evolution operator
\cite{djo-mentor}.

\section{Adelic quantum mechanics}

In foundations of standard quantum mechanics (over $\Bbb R$) one
usually starts with a representation of the canonical commutation
relation
\begin{equation}
\label{djo-com}
[\hat q,\hat k]=i\hbar ,
\end{equation}
where $q$ is a coordinate and $k$ is the
corresponding momentum. It is well known that the procedure of
quantization is not unique.
In formulation of $p$-adic quantum mechanics 
\cite{djo-vladi2,djo-ruelle} the multiplication
$\hat q\psi\rightarrow x \psi$ has no meaning for $x\in\Bbb Q_p$
and $\psi(x)\in \Bbb C$. Also, there is no
possibility to define $p$-adic "momentum" or "Hamiltonian" operator. In the real
case they are infinitesimal generators of space and time translations, but, 
since $\Bbb Q_{p}$ is disconnected field, these infinitesimal transformations become meaningless.
However, finite transformations remain meaningful and the corresponding Weyl and
evolution operators are $p$-adically well defined. For the one dimensional 
systems which classical evolution  can be described by
\begin{equation}
z_{t}=T_t z,\quad z_t= \left (
\begin{array}{c}
q(t) \\
k(t)
\end{array}
\right ), \ \ 
z= \left (
\begin{array}{c}
q(0) \\
k(0)
\end{array}
\right ),
\end{equation}
where $q(0)$ and $k(0)$, are initial position and momentum, respectively, and
$T_t$ is a matrix.
Canonical commutation relation in $p$-adic case can be represented by
the Weyl operators ($h=1$)
\begin{equation}
\hat Q_p(\alpha) \psi_p(x)=\chi_p(\alpha x)\psi_p(x)
\end{equation}
\begin{equation}
\hat K_p(\beta)\psi(x)=\psi_p(x+\beta) .
\end{equation}
Now, to the relation (\ref{djo-com}) in the real case, corresponds
\begin{equation}
\hat Q_p(\alpha)\hat K_p(\beta)=\chi_p(\alpha\beta)\hat K_p(\beta)\hat Q_p(\alpha)
\end{equation}
in the $p$-adic one. 
It is possible to introduce the family of unitary operators
\begin{equation}
\hat W_p(z)=\chi_p(-\frac 1 2 qk)\hat K_p(\beta)\hat Q_p(\alpha), \quad
z\in\Bbb Q_p\times\Bbb Q_p,
\end{equation}
that is a unitary representation of the Heisenberg-Weyl group.
Recall that this group consists of the elements $(z,\alpha)$ with the
group product
\begin{equation}
(z,\alpha)\cdot
(z',\alpha ')=(z+z',\alpha+\alpha '+\frac 1 2
B(z,z')),
\end{equation}
where $B(z,z')=-kq'+qk'$ is a skew-symmetric bilinear
form on the phase space.

Dynamics of a $p$-adic quantum model is described by a unitary
operator of evolution $U(t)$ without using the Hamiltonian. Instead of 
that, the evolution operator has been formulated in terms of its kernel
$K_t(x,y)$
\begin{equation}
U_p(t)\psi(x)=\int_{\Bbb Q_p}K_t(x,y)\psi(y) dy.
\end{equation}
The next section will be devoted to the path integral formulation and calculation 
of the quantum propagator $K_t(x,y)$ on $p$-adic spaces.

In this way \cite{djo-vladi2} $p$-adic quantum mechanics is given by
a triple
\begin{equation}
(L_2(\Bbb Q_p), W_p(z_p), U_p(t_p)).
\end{equation}
Keeping in mind that standard quantum mechanics can be also given as the 
corresponding triple, ordinary and $p$-adic 
quantum mechanics can be unified in 
the form of adelic quantum mechanics \cite{djo-mentor}
\begin{equation}
(L_2(A), W(z), U(t)).
\end{equation}
$L_{2}(\Bbb A)$ is the Hilbert space on $\Bbb A$, $W(z)$ is a unitary representation of the 
Heisenberg-Weyl group on $L_2(\Bbb A)$ and $U(t)$ is a  unitary representation of the 
evolution operator on $L_2(\Bbb A)$. 

The evolution operator $U(t)$ is defined by 
\begin{equation}
U(t)\psi(x)=\int_{\Bbb A} K_t(x,y)\psi(y)dy=\prod\limits_{v}{}
\int_{\Bbb Q_{v}}K_{t}^{(v)}(x_{v},y_{v})\psi^{(v)}(y_v) dy_{v}.
\end{equation}
The eigenvalue problem for $U(t)$ reads
\begin{equation}
U(t)\psi _{\alpha \beta} (x)=\chi (E_{\alpha} t)
\psi _{\alpha \beta} (x),
\end{equation}
where $\psi_{\alpha \beta}$ are adelic eigenfunctions, 
$E_{\alpha }=(E_{\infty}, E_{2},..., E_{p},...)$ is corresponding energy, 
indices $\alpha$ and $\beta$ denote  energy levels and their
degeneration. Note that any adelic eigenfunction has the form
\begin{equation} 
\label{djo-psi}
\Psi(x) = 
\Psi_\infty(x_\infty)\prod_{p\in S}\Psi_p(x_p)
\prod_{p\not\in S}\Omega(\mid x_p\mid_p) , \quad x\in \Bbb A, 
\end{equation}
where $\Psi_{\infty}\in L_2(\Bbb R)$,
$\Psi_{p}\in L_2(\Bbb Q_p)$.
Adelic quantum mechanics takes into account also $p$-adic quantum
effects and may be regarded as a starting point for construction of
a more complete superstring and M-theory. In the low-energy limit adelic
quantum mechanics becomes ordinary one.

\section{$p$-Adic path integrals}

A suitable way to calculate propagator in $p$-adic quantum mechanics is by
$p$-adic generalization of Feynman's path integral.
For the classical action $\bar
S(x^{\prime\prime},t^{\prime\prime};x^\prime,t^\prime)$ which is 
a polynomial quadratic in $x^{\prime\prime}$ and $x^\prime$ it is well known
that in ordinary quantum mechanics the Feynman path integral is
\begin{equation}
{\cal K}(x^{\prime\prime},t^{\prime\prime};x^\prime,t^\prime) = 
\bigg(
{i\over h}{\partial^2\bar S\over\partial x^{\prime\prime}\partial
x^\prime}\bigg)^{1/2}\exp\bigg({2\pi i\over h}\bar S
(x^{\prime\prime},t^{\prime\prime};x^\prime,t^\prime)\bigg).
\end{equation}
$p$-Adic generalization of the Feynman path integral was suggested in \cite{djo-vladi2} and 
can be written on a $p$-adic line as
\begin{equation}
\label{djo-proposal}
K_p(x^{\prime\prime}\!,\!t^{\prime\prime};\!x^\prime,\!t^\prime)=
\!\!\int
\!\!\chi_p\bigg(\!-\!{\!S[q]\over h}\bigg)\!{\cal D}q =\!\!\int\chi_p
\bigg(\!-\!{1\over h}\int^{t^{\prime\prime}}_{t^{\prime}}\!\!\!L(q,\dot
q,t)dt\bigg)\!\prod_tdq(t).
\end{equation}
In (\ref{djo-proposal}) we take $h\in \Bbb Q$ and $q,t\in \Bbb Q_p$.
This path integral is elaborated, for the first time, for the
harmonic oscillator \cite{djo-zele}. It was shown that there exists the limit
\begin{eqnarray}
K_p(x^{\prime\prime},t^{\prime\prime};x^\prime,t^\prime)=
\lim_{n\to\infty}
K_p^{(n)}(x^{\prime\prime},t^{\prime\prime};x^\prime,t^\prime) =
\lim_{n\to\infty}N^{(n)}_p(t^{\prime\prime},t^\prime) \nonumber \\
\times\int_{\Bbb Q_{p}}\cdots
\int_{{\Bbb Q}_{p}}
\chi_p\bigg(-{1\over h}\sum^n_{i=1}\bar S(
q_i,t_i;q_{i-1},t_{i-1})\bigg)dq_1\cdots dq_{n-1}\ ,
\end{eqnarray}
where $N^{(n)}_p(t^{\prime\prime},t^\prime)$ is the corresponding
normalization factor for the harmonic oscillator.
The subdivision of $p$-adic time segment $t_0<t_1<\cdots<t_{n-1}<t_n$ 
is made 
according to linear order on $\Bbb Q_p$ and  $\mid
t_i-t_{i-1}\mid_v\to0$ for every $i = 1,2,\cdots,n$, when
$n\to\infty$. In the similar way we have calculated
path integrals for: a particle in a constant external field
\cite{djo-pristina},
some minisuperspace cosmological models and a relativistic free
particle \cite{djo-mentormi}, as well as for a harmonic oscillator with a 
time-dependent frequency \cite{djo-pristina}.

$p$-Adic classical mechanics  has the same analytic form as in 
the real
case. If $q(t)=\bar q(t)+y(t)$ denotes a possible
quantum path, with conditions $y(t^\prime) = y(t^{\prime\prime}) = 
0$, where
$\bar q(t)$ is a $p$-adic classical path with $\delta S[\bar q] = 0$, 
we have the following action for quadratic Lagrangians:
\begin{equation}
\label{djo-razvoj}
S[q] = S[\bar q]+{1\over2!}\delta^2S[\bar q] = S[\bar q]+{1\over2}
\int^{t^{\prime\prime}}_{t^{\prime}}\bigg(y{\partial\over\partial q}
+\dot y{\partial\over\partial\dot q}\bigg)^{(2)}L(q,\dot q,t)dt.
\end{equation}
Putting (\ref{djo-razvoj}) into (\ref{djo-proposal}), and using condition 
\begin{equation}
\int_{\Bbb Q_p}
\!K_p^*(x^{\prime\prime}\!,t^{\prime\prime}\!;x^\prime,t^\prime)
K_p(z,t^{\prime\prime}\!;x^\prime\!,t^\prime)dx^\prime =
\delta_p(x^{\prime\prime}-z),
\end{equation}
with quadratic expansion of action
as well as the general form of the normalization
factor 
$$
N_p(t^{\prime\prime},t^\prime) = \mid 
N_p(t^{\prime\prime},t^\prime)\vert_\infty 
A_p(t^{\prime\prime},t^\prime ),
$$
we obtain general expression for the propagator (for some details,  see  \cite{djo-general})
\begin{equation}
\label{djo-quadratic}
K_p(x^{\prime\prime}\!,t^{\prime\prime}\!;x^\prime\!,\!t^\prime)= 
\!\lambda_p
\!\bigg(\!-\frac{1}{2h}{\partial^2\bar S\over \partial
x^{\prime\prime}\partial x^\prime}
\bigg)\!
\Big\arrowvert\frac{1}{h}{\partial^2\bar S\over \partial
x^{\prime\prime}\partial x^\prime}
\Big\arrowvert^{\frac{1}{2}}_p
\chi_p
\!\bigg(\!-{1\over h}\bar 
S(x^{\prime\prime}\!,t^{\prime\prime}\!;x^\prime\!,\!t^\prime)
\bigg)\!.
\end{equation}
This result exhibits some very important 
properties. For instance, replacing an index $p$ with $v$ 
in (\ref{djo-quadratic}) we can write quantum-mechanical amplitude $K$ in ordinary and 
all $p$-adic 
cases in the same compact form. It points out a generic behaviour of 
quantum propagation
in Archimedean  and non-Archimedean spaces and emphasizes the fundamental role 
of the Feynman path
integral method in quantum theory. Also, considering the most general 
quadratic
$p$-adic Lagrangian $L(x,\dot x,t)=a(t)\dot x^2+2b(t)\dot 
xx+c(t)x^2+
2d(t)\dot x+2e(t)x+f(t)$ with analytic coefficients, we found 
a connection \cite{djo-veza}
between these coefficients and  the simplest 
$p$-adic quantum state
$\Omega (|x|_p)$, that is necessary for existence of adelic quantum dynamics.
For space-time discreteness in adelic models, see \cite{djo-mentormi}.

It is worth mentioning  that this approach can be extended to systems with the two,
three and more dimensions, and results will be presented elsewhere.
The above  results are also a starting point for a further elaboration
of adelic quantum mechanics and for a semiclassical
computation
of the $p$-adic path integrals with non-quadratic  Lagrangians.

\section{Adelic quantum cosmology}

Adelic quantum cosmology \cite{djo-dn} is an application of adelic
quantum mechanics to the universe as a whole. It unifies ordinary and
$p$-adic quantum cosmology. Here, path integral formalism occurs to be
quite appropriate tool to take integration over both Archimedean and non-Archimedean
geometries on the equal footing. In this approach we introduce $\upsilon$-adic
complex-valued cosmological amplitudes by a functional integral
\begin{equation}
\langle
h_{ij}^{\prime\prime},\phi^{\prime\prime},\Sigma^{\prime\prime}|
h_{ij}^\prime,\phi^\prime,\Sigma^\prime\rangle_\upsilon=
\int{\cal D}{(g_{\mu\nu})}_\upsilon{\cal D}(\Phi)_\upsilon
\chi_\upsilon(-S_\upsilon[g_{\mu\nu},\!\!\Phi]).
\end{equation}
In practice, it is not possible to deal with full superspace
(the space of all 3-metrics and matter field configurations).
Instead, one exploits minisuperspace (a finite number of
coordinates $(h_{ij},\phi)$). After this simplification, 
$\upsilon$-factors of adelic
minisuperspace propagator are given by the relation
\begin{equation}
\langle {q^\alpha}^{\prime\prime}|{q^\alpha}^\prime\rangle_\upsilon
=\int dN K_\upsilon({q^\alpha}^{\prime\prime},N|{q^\alpha}^\prime,0),
\end{equation}
where $K_\upsilon$ is an ordinary quantum-mechanical propagator with fixed
minisuperspace coordinates ${q^\alpha}$ and the lapse function $N$.

We illustrate adelic quantum cosmology by Bianchi I model $(k=0)$.
Using Lorentz metric \cite{djo-ueda}
\begin{equation}
ds^2=\sigma^2\left[-\frac{N^2(t)}{a^2(t)}dt^2 
+ a^2(t)dx^2 +b^2(t)dy^2+c^2(t)dz^2\right]
\end{equation}
and  replacements:
\begin{equation}
x=\frac{bc+a^2}{2},\enskip y=\frac{bc-a^2}{2},\enskip \dot z^2=a^2\dot
b\dot c,
\end{equation}
we obtain the corresponding action
\begin{equation}
S_p[x,y,z]=\frac{1}{2}\int^1_0dt
\left[-\frac{1}{N}\left(\frac{\dot x^2-\dot y^2}{2}+\dot z^2\right)
-\lambda N(x+y)\right],
\end{equation}
and equations of motion
\begin{equation}
\ddot x+\lambda N^2=0,\quad\ddot y-\lambda N^2=0,\quad\ddot z=0.
\end{equation}
Taking into account conditions $x(0)=x',\ y(0)=y',\ z(0)=z'$,
$x(1)=x'',\ y(1)=y'',\ z(1)=z''$, the quantum transition amplitude
can be written as
\begin{equation}
\!{\cal K}_p(x'',y'',z'',N|x',y',z',0)=
\frac{\lambda_p(-2N)}{\left|4^{\frac 1 3}N\right|_p^{\frac 3 2}}
\chi_p\left(-\bar S(x'',y'',z'',N|x',y',z',0)\!
\right).
\end{equation}

 Conditions for the existence of the vacuum state 
$\Omega(|x|_p)\Omega(|y|_p)\Omega(|z|_p)$ can be calculated from
the equality
$$
\int_{|x'|_p\leq 1}\int_{|y'|_p\leq 1}
\int_{|z'|_p\leq 1}
{\cal K}_p(x'',y'',z'',N|x',y',z',0)dx'dy'dz'
$$
$$
=
\Omega(|x''|_p)\Omega(|y''|_p)\Omega(|z''|_p),
$$
and the simplest vacuum state is
\begin{equation}
\Psi_p(x,y,z,N)=\left \{
\begin{array}{ll} 
\Omega(|x|_p)\Omega(|y|_p)\Omega(|z|_p),& 
\quad |N|_p\leq1,\ |\lambda|_p\leq 1, \ p\neq 2, \\
\Omega(|x|_2)\Omega(|y|_2)\Omega(|z|_2),& \quad |N|_2\leq\frac{1}{2},
\  |\lambda|_2\leq 2,\ p=2.  
\end{array} \right.
\end{equation}

According to (\ref{djo-psi}) adelic wave function $\Psi(x,t)$ offers more
information on a physical system than only its standard part
$\Psi_\infty(x,t)$. In quantum-mechanical experiments, as well as in all measurements,
numerical results belong to the field of rational numbers $\Bbb Q$.
For the  Bianchi I model, as well as for any adelic quantum model, according to the usual
interpretation of the wave function we have to consider 
$\vert \Psi(x,t)\vert^2_\infty $ at rational space-time  points.  In the
above adelic case we get 
$$
|\Psi(x,y,z,N)|^2_\infty =
|\Psi_\infty(x,y,z,N)|^2_\infty \prod_p \Omega(| x |_p)\Omega(| y
|_p\Omega(| z |_p)
$$
\begin{equation}
=\left \{
\begin{array}{ll}
|\Psi_\infty(x,y,z,N)|^2_\infty\ , & x,y,z \in\Bbb Z\ , \\ 
0\ , & x,y,z \in \Bbb Q\setminus \Bbb Z .
\end{array}\right .
\end{equation}
Here we used the following properties
of the $\Omega$-function: $\Omega^2(\vert x\vert_p) = \Omega(\vert
x\vert_p)$, $\prod_p\Omega(\vert x\vert_p) =1$ if
$x\in\Bbb Z$, and $\prod_p\Omega(\vert x\vert_p) =0$ if
$x\in\Bbb Q\setminus\Bbb Z$. Thus, it means that
positions $x,y,z$ may have only discrete values: $x =0,\pm 1,\pm
2,...$. Since  the $\Omega$-function is invariant under the Fourier
transformation, there is also discrete momentum space. 
When system is in some excited state, the sharpness of the discrete structure
disappears and space demonstrates usual continuous properties. 
It is worth mentioning that a space-time discreteness is also noted in the framework
of q-deformed quantum mechanics \cite{djo-wess}.

\section{$p$-Adic analysis and q-analysis.  The Moyal product}

Some connections between  $p$-adic analysis and quantum
deformations has been noticed \cite{djo-f} in a variety of cases during
the last ten years or so. It was shown \cite{djo-fz}  that the two 
parameter Sklyanin quantum 
algebra  and its generalizations provide a promising connection between
the $p$-adics  and quantum deformation. A similar 
connection has been  indicated by Macdonald's paper 
\cite{djo-mac} on orthogonal polynomials associated with the root
systems. In \cite{djo-fz} it was also pointed out that elliptic
quantum group and its generalizations unify the $p$-adic and real
versions of a Lie group (e.g. $SL(2$)). This result is 
connected with adelic approach and the possibility of establishing
q-deformed Euler products. 

In some other  contexts it has been observed that the Haar measure on
$SU_q(2)$ coincides with the Haar measure on the field of $p$-adic
numbers $\Bbb Q_p$ if $q=\frac{1}{p}$ \cite{djo-av}.
Namely, Tomea-Jackson
integral in q-analysis
\begin{equation}
\int^1_0 f(x)d_qx=(1-q)\sum_{n=0}^\infty f(q^n)q^n,
\end{equation}
and the integral in $p$-adic analysis
\begin{equation}
\int_{|x|_p\le 1}  f(|x|_p)dx=(1-\frac 1 p)\sum_{n=0}^\infty f(p^{-n})p^{-n},
\end{equation}
are equal if $q=\frac 1 p$, {\it i.e.}
\begin{equation}
\int^1_0 f(x)d_{1/p}x=\int_{|x|_p\le 1}  f(|x|_p)dx.
\end{equation}

In q-analysis there is the following differential operator 
(related to the q-deformed momentum in the coordinate
representation \cite{djo-av})
\begin{equation}
\label{djo-star}
\partial_q f(x)=\frac{f(x)-f(qx)}{(1-q)x}.
\end{equation}
In $p$-adic analysis, when one considers a complex-valued
function $f(x)$ depending on a $p$-adic variable $x$ we are not 
able to use standard definition of differentiation. Instead of
that it is possible to use Vladimirov's operator
\begin{equation}
\label{djo-twostar}
D^{\alpha}\psi (x)= \frac{p-1}{1-p^{-1-\alpha}}
\int \frac{f(x)-f(y)}{|x-y|_p^{\alpha+1}} dy
\end{equation}
which in a sense resembles (\ref{djo-star}).
Moreover, there is a potential such that the spectrum of the $p$-adic Schr\"odinger- like 
( diffusion ) equation
\cite{djo-vvsch} 
\begin{equation}
\label{djo-povisilica}
D\psi (x)+V(|x|_p)\psi (x)=E\psi (x)
\end{equation}
is the same one as in the case of $q$-deformed oscillator found by
Biedenharn \cite{djo-b} and Macfarlane \cite{djo-m} for
$q=1/p$. For more details, see \cite{djo-av}.

Recently
\cite{djo-trnovo}, it has been proposed a new pseudodifferential operator
with rational part of $p$-adic numbers $\{x\}_p$.
In such case,  energy levels for $p$-adic free particle
exhibit discrete dependence on the corresponding momentum:
$\{E\}_p=\{k\}_p^2$. Note also a proposal for
q-deformation of Vladimirov's operator \cite{djo-kozyrev}.

We see that there are some interesting relations between $p$-adic and
q-analysis, and in a sense between adelic quantum mechanics and 
noncommutative one. It would be fruitful to find some deeper reasons for these 
connections, between theories which
pretend to give us  more insights on the space-time structure at the
Planck scale. By now it is not enough understood. It seems to be 
reasonable to formulate 
a noncommutative adelic quantum mechanics that may  connect
non-Archimedean and noncommutative effects and structures. As
the first step in this direction one has to consider 
a $p$-adic and adelic generalization of the Moyal product.

Let us consider D-dimensional classical space with coordinates
$x^1,x^2$, $\cdots,x^D$. Let $f(x)$ be a classical function
$f(x)=f(x^1,x^2,\cdots,x^D)$.
Then, with the respect to the Fourier transformations, we have 
\begin{equation}
\tilde f(k)=\int_{\Bbb Q_\upsilon^D} dx \ \chi_v(kx) f(x),
\end{equation} 
\begin{equation}
f(x)=\int_{\Bbb Q_\upsilon^D} dk \ \chi_v(-kx) \tilde f(k).
\end{equation}
According to the usual Weyl quantization
\begin{equation}
\hat f(x)=\int_{\Bbb Q_\infty^D} dk \ 
\chi_\infty(-k \hat x) \tilde f(k)\equiv  f(\hat x).
\end{equation}
Let us now have two classical functions $f(x)$ and $g(x)$ with
\begin{equation}
\hat f(x)=\int_{\Bbb Q_\infty^D} dk \ \chi_\infty(-k \hat x) \tilde f(k),
\end{equation}
\begin{equation}
\hat g(x)=\int_{\Bbb Q_\infty^D} dk \ \chi_\infty(-k \hat x) \tilde g(k).
\end{equation}
In the coordinate representation we can write the same above expressions replacing
$\hat x $  by  $ x$ and extend it to all $p$-adic cases.

Now we are interested in product $\hat f(x) \hat g(x)$. In the real case this
operator product is of the form
\begin{equation}
(\hat f \cdot \hat g)(x)=\int \int dk dk' \ \chi_\infty(-k\hat x)
\chi_\infty(-k' \hat x) \tilde f(k)\tilde g(k').
\end{equation}
Using the Baker-Campbell-Hausdorff formula, the relation (\ref{djo-noncom}) 
and then the coordinate representation one finds
the  Moyal product in the form
\begin{equation}
(f\ast g)(x)=\int\int dk dk' \ \chi_\upsilon\left ( -(k+k')x+\frac 1 2
k_ik'_j\theta^{ij}\right )\tilde f(k)\tilde g(k'),
\end{equation}
where we already used our generalization from  $\Bbb Q_\infty$ to $\Bbb Q_\upsilon$. 
Note that in the real case we  use 
$k_i\rightarrow -(i/2\pi)(\partial/\partial x^i)$ and obtain the well
known form
\begin{equation}
(f\ast g)(x)=\chi_\infty\left(-\frac{\theta^{ij}}{2(2\pi)^2}
\frac{\partial}{\partial y^i}\frac{\partial}{\partial z^j}\right ) f(y)g(z)|_{y=z=x}.
\end{equation}
Thus, as the $p$-adic Moyal product we take
\begin{equation}
(\hat f \ast \hat g)(x)=\int_{\Bbb Q_p^D}\int_{\Bbb Q_p^D} dk dk'
\ \chi_p(-(x^ik_i+x^jk'_j)+\frac{1}{2} k_ik'_j\theta^{ij})\tilde f(k)\tilde g(k').
\end{equation}
As the first step in adelization one can consider the Moyal product  on
$\Bbb R\times \prod_{p\in S} \Bbb Q_p$ $\times \prod_{p\not\in S}\Bbb Z_p$ space.
Various adelic aspects of the Moyal product will be presented elsewhere.

\bigskip
{\bf{Acknowledgments}} Authors G.Dj. and B.D. wish to thank the
co-directors of ARW  "Noncommutative Structures in Mathematics and Physics" 
Profs. J. Wess and S. Duplij for their invitation to participate
and give a talk. G.Dj. is partially supported by DFG Project
``Noncommutative space-time structure - Cooperation with Balkan Countries''.
The work of B.D. was supported in part by RFFI grant 990100866. 
\end{article}

\end{document}